# A study of the optical and polarisation properties of InGaN/GaN multiple quantum wells grown on a-plane and m-plane GaN substrates

D. Kundys[1,4], D. Sutherland[1], M. Davies[1], F. Oehler[2], J. Griffiths[2], P. Dawson[1], M.J. Kappers[2], C.J. Humphreys[2], S. Schulz[3], F. Tang[2] and R.A. Oliver[2]

[1]*School of Physics and Astronomy, Photon Science Institute, University of Manchester, Manchester, M13 9PL, UK*
[2]*Department of Materials Science and Metallurgy, University of Cambridge, CB3 0FS, UK*
[3]*Tyndall National Institute, Dyke Parade, Cork, Ireland*
[4]*School of Engineering and Physical Sciences, Heriot-Watt University, Edinburgh EH14 4AS, UK*

In this paper we report on a comparative study of the low temperature emission and polarisation properties of InGaN/GaN quantum wells (QWs) grown on nonpolar ($11\bar{2}0$) *a*-plane and ($10\bar{1}0$) *m*-plane free-standing bulk GaN substrates where the In content varied from 0.14 to 0.28 in the *m*-plane series and 0.08 to 0.21 for the *a*-plane series. The low temperature photoluminescence spectra from both sets of samples are very broad with full width at half-maximum height increasing from 81 to 330 meV as the In fraction increases. Comparative photoluminescence excitation spectroscopy indicates that the recombination mainly involves strongly localised carriers. At a temperature of 10 K the degree of linear polarisation of the *a*-plane samples is much smaller than of the *m*-plane counterparts and also varies across the spectrum. From polarisation-resolved photoluminescence excitation spectroscopy we measured the energy splitting between the lowest valence sub-band states to lie in the range of 23-54 meV for both *a*-and *m*-plane samples in which we could observe distinct exciton features in the polarised photoluminescence excitation spectroscopy. Thus, the thermal occupation of a higher valence subband cannot be responsible for the reduction of the degree of linear polarisation. Time-resolved spectroscopy indicates that in *a*-plane samples there is an extra emission component which at least partly responsible for the reduction in the degree of linear polarisation.

*PACS numbers: 78.55.Cr, 78.67.De, 78.47.jd*

## Introduction

Nonpolar InGaN/GaN multiple quantum wells (MQWs) are attracting a great deal of interest because of their ability to emit linearly polarised light. This characteristic may be of practical use in, e.g. back-lit liquid crystal displays [1]. The emission of polarized light occurs because of the breaking of the crystal symmetry due to anisotropic in-plane strain [2] [3]. The degree of linear polarisation (DLP) of the total light emission ideally should depend on the relative populations of the two lowest energy valence subbands. The DLP of the emission is defined (in terms of the geometry of our experiment, see figure 1) as follows:

$$\text{DLP} = \frac{(I_\perp - I_\parallel)}{(I_\perp + I_\parallel)} \quad (1)$$

For any application that relies on polarised light, it is not only vital that the DLP be high but also the efficiency of the intrinsic emission be as large as possible. This latter issue can be affected by the choice of substrate.

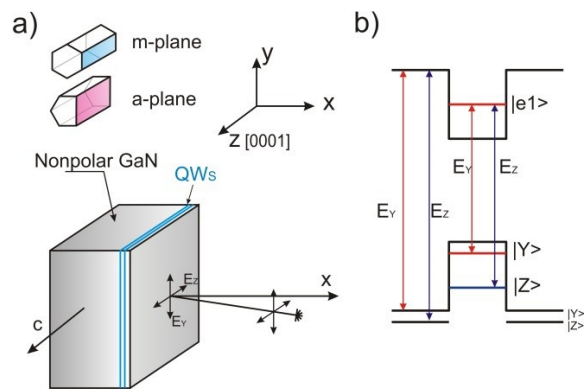

FIG 1. Experimental setup geometry is illustrated in **a)** where the light excitation and collection is close to normal incidence and along the x-axis. The QW energy diagram is shown in **b)** with relative interband transitions illustrated for both QW and GaN barriers associated with |Y⟩ and |Z⟩ subbands. The transitions associated with |X⟩ state are not allowed because its dipole moment is equal to zero due to experimental geometry configuration.

Various substrates are available for heteroepitaxial growth of nonpolar GaN such as SiC [4], LiAlO$_3$ [5], and sapphire [6] [7] [8] [9]. Heteroepitaxial growth, however, results in high densities of extended defects, despite the use of sophisticated defect reduction techniques [10] [11]. Typically the densities of threading dislocations and basal plane stacking faults for heteroepitaxial non-polar growth are $(2-7) \times 10^9$ cm$^{-2}$ and $\sim 10^5$ cm$^{-1}$ respectively [12]. Such high densities of defects can lead to extrinsic non-radiative and radiative processes that reduce the efficiency of the intrinsic emission process. For these reasons the most promising choice for high efficiency, large DLP structures is homoepitaxial growth on freestanding, ammonothermal GaN substrates with a threading dislocation density lower than $10^5$ cm$^{-2}$ and negligible basal plane stacking faults.

At the moment InGaN/GaN QWs grown on *m*-plane GaN repeatedly show superior polarization properties with DLP values over 90% at room temperature [13] compared with structures grown on *a*-plane GaN [14] [15]. From a theoretical point of view [16] the change of nonpolar plane should make no difference to the polarization properties of InGaN/GaN QWs. At the moment this behavior is not understood. In this paper we report on a study of the optical and polarisation properties InGaN/GaN quantum well QW grown on ammonothermal *a*- and *m*-plane GaN substrates. We have used a combination of polarized photoluminescence (PL) spectroscopy, time resolved spectroscopy and polarization resolved photoluminescence excitation (P-PLE) spectroscopy to compare the properties of the QWs where the differences between the sample sets are entirely due to the growth on the different non-polar planes.

**Experimental details**

All the samples were grown by metalorganic vapour phase epitaxy in a 6x2″ Thomas Swan close-coupled showerhead reactor. The InGaN/GaN MQW samples were grown on nonpolar ammonothermal GaN substrates with intentional crystal misorientations for the $(11\bar{2}0)$ *a*-plane of 0±0.2° along the *m*-axis and 0.3±0.2° towards the +*c*-axis, and for $(10\bar{1}0)$ *m*-plane 0±0.2° along the *a*-axis and 2.0±0.2° towards the (-*c*)-axis. Trimethylgallium (TMG), trimethylindium (TMI) and ammonia (NH$_3$) were used as the precursors with hydrogen as the carrier gas for the growth of the GaN epilayer and nitrogen for the growth of InGaN QWs and GaN barrier layers. First, a GaN buffer layer with a thickness of 800 to 1000 nm was grown at 1050°C followed by the InGaN/GaN QW stacks, which were grown using a "quasi-two temperature" growth method, adapted from a method commonly used for growth on *c*-plane [17]. The In composition of the samples was varied by changes in the growth temperature from 690°C to 705 °C for the *a*-plane samples and 705°C to 745°C for the *m*-plane samples (see Table I). Following the growth of the InGaN QWs a 1.0 nm GaN cap was grown at the InGaN growth temperature. The TMG and NH$_3$ flows were then maintained during a 90 s temperature ramp to 860 °C, at which point the barrier growth was completed. Every sample was analysed using a high resolution X-ray diffraction (XRD) MRD diffractometer from Panalytical, equipped with a symmetric 4-bounce monochromator and a 3-bounce analyser to select the CuK$\alpha_1$ wavelength.

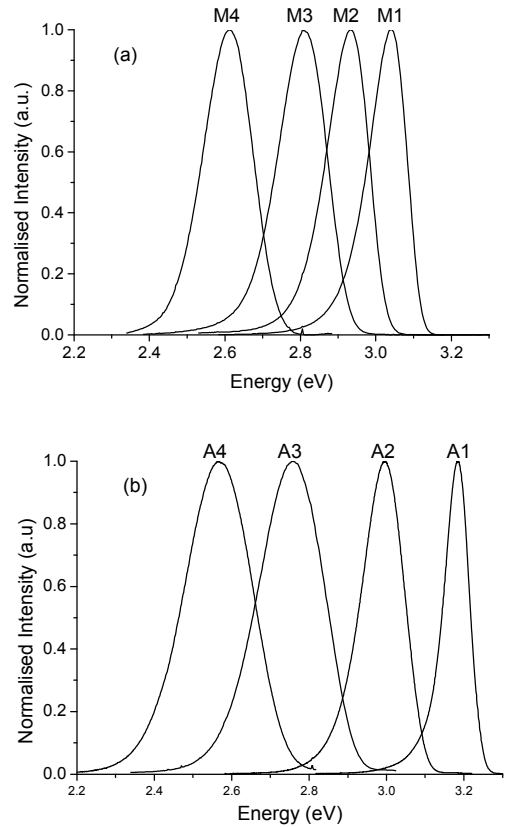

Figure 2. Photoluminescence spectra measured at 10K are shown in (a) and (b) for *m*-plane and *a*-plane samples respectively identified by the numbers in Table I.

| Sample Number | Growth temperature of (°C) | In fraction in the QWs | Thickness of QWs (nm) | GaN barrier thickness | Peak emission energy (eV) | FWHM of emission spectrum | Stokes shift (meV) | Lowest energy exciton splitting (meV) |
|---|---|---|---|---|---|---|---|---|
| M1 | 745 | 0.14 | 1.9 | 6.1 | 3.038 | 116 | 131 ±2 | 38 ± 2 |
| M2 | 735 | 0.18 | 2.0 | 6.1 | 2.928 | 135 | 180 ±3 | 35 ± 4 |
| M3 | 725 | 0.22 | 2.1 | 6.1 | 2.802 | 160 | 264 ±4 | 54 ± 5 |
| M4 | 705 | 0.28 | 2.4 | 6.1 | 2.610 | 157 | - | - |
| A1 | 705 | 0.08 | 2.0 | 6.0 | 3.194 | 81 | 87± 2 | 23 ± 2 |
| A2 | 695 | 0.13 | 2.1 | 6.0 | 2.980 | 136 | 137 ±2 | 38 ± 2 |
| A3 | / | 0.16* | 2.25* | 6.0 | 2.765 | 208 | 276 ±4 | 48 ± 5 |
| A4 | 690 | 0.21 | 2.4 | 6.0 | 2.527 | 330 | - | - |

Table I: *m*-plane (A series) and *a*-plane (M series) growth temperature and sample properties obtained from XRD and TEM along with optical data measured at 10K.

The well and barrier widths and QW compositions (as detailed in Table I) were determined using ω-2θ scans of the brightest symmetric reflection with a range of 10° in ω following the approach of Vickers *et al.* [18] and modified for nonpolar orientations [19]. The missing satellite peaks typically present in XRD scans of MQW structures used to determine the well widths were absent for these samples. Hence, high angle annular dark field scanning transmission electron microscopy (HAADF-STEM) was used to measure the well and barrier widths of samples A2, A4, and M2 to inform the calculation of the layer thicknesses in the remaining samples. HAADF-STEM was performed with an FEI Titan fitted with both probe and image aberration correctors operated at 300 keV. The results of the structural analysis are shown in Table I.

The PL and P-PLE studies were carried out either using excitation from a CW He/Cd laser or a combination of a 300W Xenon lamp and monochromator as a fixed or tuneable wavelength excitation source. The samples were mounted in the cryostat so that *c*-axis of the GaN was horizontal.

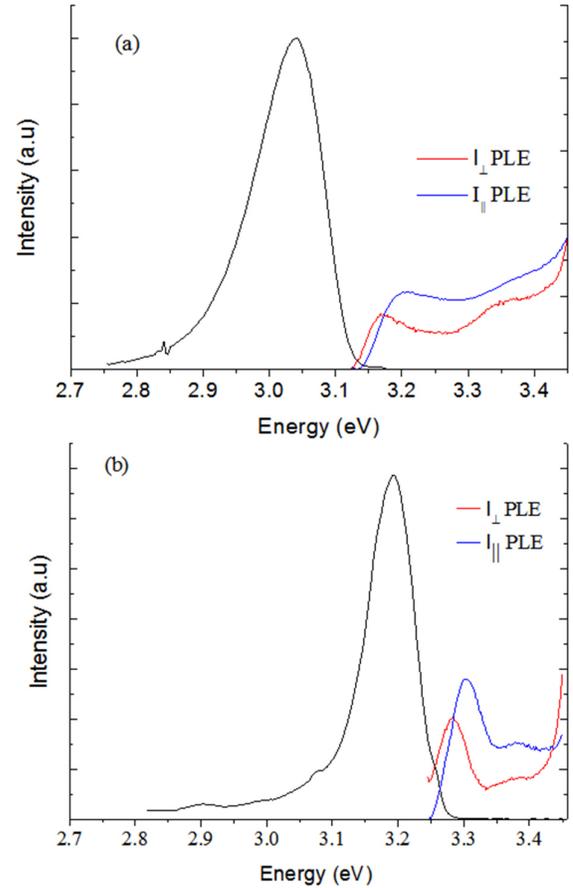

Figure 3. Low temperature (10K) PL spectra (black line) with PLE spectra are shown in (a) and (b) for *m*- and *a*-plane samples M1 and A1, respectively for two different linear polarisations of excitation radiation.

The PL from the samples was analysed by a 0.85 m double grating spectrometer and a Peltier-cooled GaAs photomultiplier using standard lock-in detection techniques. The spectral dependence of the degree of polarisation (DLP(E)) of the emission was determined by measuring the PL spectra polarised parallel ($I_\parallel$) and perpendicular ($I_\perp$) to the *c*-axis of the sample in question and then combining the polarised spectra as described by equation 1.

The polarisation dependent excitation spectroscopy reveal exciton transitions associated with the different valence subbands; this not only enables measurement of the valence sub-band energy splitting as described previously [20] [3] [21] [22] [23] but can also be used to help to identify the nature of the recombination processes.

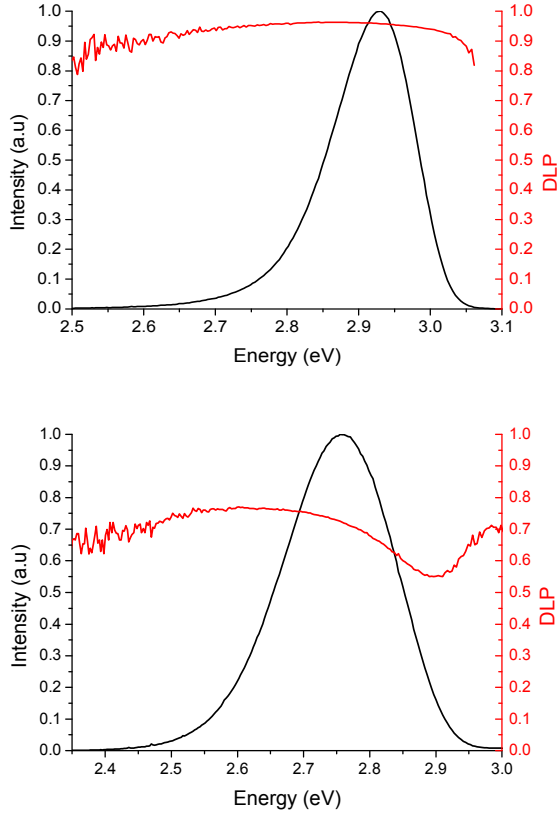
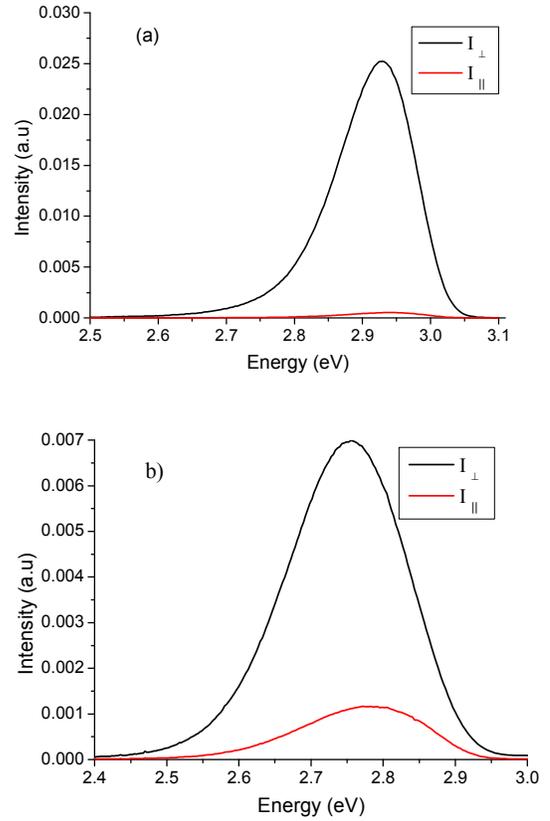

Figure 4. Photoluminescence spectra (black solid lines) and corresponding spectral dependence of the DLP(E) (red dashed line) measured at 10K are shown in top and bottom for *m*-plane and *a*-plane samples M3 and A3 respectively

Figure 5. PL spectra for (a) M2 and (b) A3 detected when monitoring light with polarisations in the $I_\perp$ and $I_\parallel$ directions as indicated.

For the time-resolved PL and PL decay time measurements, a frequency doubled mode-locked Ti:sapphire laser operating at a repetition rate of 800 kHz and emitting light with a photon energy of 3.18 eV was used as the excitation source and the signal processed with a time-correlated single photon counting system.

**Results and Discussion**

In Figures 2(a) and 2(b) low temperature (T = 10 K) photoluminescence spectra are shown for the InGaN/GaN MQW samples grown on *m*- and *a*- planes respectively. The most noticeable aspect of all the PL spectra is the particularly large line widths which, depending on the In fraction, lie in the range 81 to 330 meV (see Table I). These line widths are much larger than the equivalent InGaN/GaN QWs grown on *c*-plane substrates, e.g. a polar InGaN/GaN single QW structure with an In fraction of 0.25 exhibited [24] a low temperature PL linewidth of 82 meV whereas A4 and M2 (samples with similar In fractions) have linewidths of 330 and 160 meV respectively. One cause of the large linewidths could be inter-well variations in the In fraction and/or quantum well thickness. However for the *m*- plane samples this is certainly not the case as the linewidth of a single QW structure grown under the same conditions as a MQW structure had similar spectroscopic characteristics.

A comparison of the P-PLE spectra and the PL spectra throws light on the nature of the dominant recombination process. We were only able to observe distinct exciton transitions associated with the two lowest valence subbands in samples M1, M2, M3, A1, A2 and A3. For the samples A4 and M4 the spectra were inhomogeneously broadened to such an extent that discrete peaks could not be resolved. The P-PLE spectra for samples M1 and A1 are shown in Figure 3. We attribute the peaks that lie between 3.05eV and 3.35 eV to exciton transitions involving the n = 1 confined electron states and the n = 1 confined strain split valence bands. The most striking aspect of the data presented in Figure 3 is the very large energy difference (180 meV) between the energies of the lowest energy QW exciton transition and the peak energy of the PL emission. This energy difference is often referred to as the Stokes shift. Values for the Stokes shift for the samples in which exciton transitions could be resolved are listed in Table I. Previously the emission at low temperatures from non-polar InGaN/GaN QWs has been attributed to localised exciton recombination [25]. On this basis the values of the Stokes shifts suggest very large exciton localisation energies that increase with increasing In fraction. It is also noticeable that the line width of the exciton features in the P-PLE spectrum are much smaller than the PL line width, e.g. in the case of the data presented in Figure 3 the linewidth of the lowest energy exciton peak is 60 meV compared with a PL linewidth of 130 meV. One possible interpretation for this observation is as follows: the exciton transitions in the P-PLE spectrum are associated with the creation of excitons that are free in the plane of the QWs. If the length scale of the localised disorder is less than the in-plane exciton Bohr radius so that, to an extent, in the PLE spectra the effects of the disorder are averaged out. Yet when the carriers become localised the localisation itself introduces an energy variation at the different localisation sites, perhaps due to variations in quantum confinement in the plane of the QWs.

The nature of the localisation is currently under investigation, but as in polar *c*-plane InGaN/GaN QWs [26] [27], it is probable that the effects of the local concentration of In atoms plays a dominant role. In the *c*-plane case these local variations in indium content are mainly statistical fluctuations in a random alloy and indeed recent theoretical modelling [28] has indicated that in *m*-plane QWs hole localisation at random In fluctuations is largely responsible for the low temperature PL linewidth. The major difference between polar and non-polar QWs of course is the absence of an electric field perpendicular to the plane of the QWs, an effect that strongly influences the radiative recombination dynamics in polar structures [29]. However, as demonstrated recently [28] the absence of the electric field has significant consequences for the electron localisation. In polar structures the electric field results in the electrons being localised largely by the well width fluctuations, while in non-polar structures this effect is greatly reduced so that the electrons become bound to the localised holes resulting in localised excitons.

Also of importance is the fact that the P-PLE data reveals directly (assuming no difference in exciton binding energies for the different valence bands) the energy difference of the two lowest lying valence bands. For the samples in which we were able to determine the exciton splitting (listed in Table I) the values are $23 \pm 2$ meV (A1), $35 \pm 5$ meV (A2), $48 \pm 5$ meV (A3), $38 \pm 2$ meV (M1), $35 \pm 4$ meV (M2) and $54 \pm 5$ meV (M3). These values are similar to those reported elsewhere [13] [21] [22] [30] from polarised PL and electroluminescence measurements on InGaN QW structures. Although it should be born in mind that a direct comparison of the values reported here and those in the cited works is not strictly valid. This is because the data reported here essentially reflects the differences between the free exciton transitions associated with the different valence subbands whereas the referenced data involves measurements involving recombination that could well be influenced by the effects of carrier localisation.

Shown in Figure 4 are the spectral dependences of the DLP(E) for the *a*- and *m*-plane samples A3 and M2. The contrasting spectra are representative of the behaviour from the two sample sets. The *m*- plane sample exhibits a high value of DLP ($> 0.9$) which is constant across the spectrum, but the *a*- plane sample exhibits a maximum value of the DLP $\sim 0.6$ and the DLP(E) spectrum is not flat and exhibits a pronounced reduction on the high energy side of the spectrum. The first question to ask is whether there is any evidence for a difference in the lowest energy valence subband splitting in the *m*- and *a*-plane structures which could influence the DLP in the different structures. As described above the values of the valence subband splitting in the *a*- and *m*-plane samples in which we were able to observe any exciton transitions are $\geq 23$ meV.

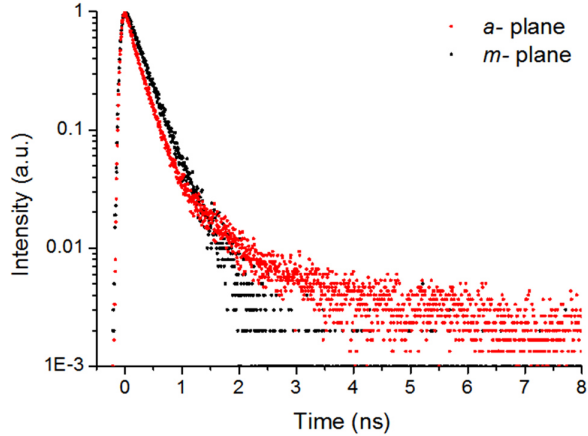

Figure 6. PL time decay curves for samples M2 and A3 as indicated in the figure.

So ignoring any effects due to localisation we conclude at a temperature of 10K that the higher lying valence states are not significantly occupied for both *a*-and *m*-plane samples. Hence the differences in the degree of DLP and the spectral dependence cannot be explained by differences in the splitting of the valence subband energies. As implied above this argument ignores the nature of the recombination and really would only valid for the case of free carrier recombination. As we have already noted this is not the case at a temperature of 10K with the recombination of localised excitons likely to be dominant. As to the dip in the DLP(E) for the *a*-plane samples if we examine the PL spectra for the two polarisations as shown in Figure 5 there is a clear difference in the peak position of the two spectra. This in itself does not necessarily suggest that there are two distinct recombination channels for the *a*-plane samples; in fact it merely represents how the DLP changes across the spectrum. We can also plot the DLP data in the form shown in Figure 5 where the spectra for both polarisations are plotted separately for an *a*-plane and a *m*-plane structure. These plots reveal that the mechanism responsible for both the low DLP and the dip in the DLP(E) spectrum of the *a*-plane sample is common to both the *a*- and *m*-plane structures, with whatever the cause of this behaviour being much less important in the *m*-plane samples. To throw further light on the differences in the spectral dependence of the polarisation data of the *a*- and *m*-plane samples we performed PL time decay measurements and time-resolved spectroscopy. The time decay data measured at the peak of the spectra for samples M2 and A3 are shown in Figure 6. All the *m*-plane samples exhibit single exponential decays with time constants ~300 ps which remain unchanged across the individual spectra. This is similar to the data that has already been reported elsewhere [25]. However, the PL decay curves from the *a*-plane samples exhibit an additional component that decays over a much longer time scale. It should also be noted that the time constant of the rapidly decaying component from the *a*-plane sample is very similar to that measured on the *m*-plane sample, implying that the majority of the emission from the *a*- and *m*-plane samples involves an identical recombination process. Also the longer lived component increases in relative strength as the detection energy is moved to the high energy side of the spectrum. The results of the time resolved spectroscopy shown in Figure 7 reflect this observation. In the time window 0-1 ns where the decay is dominated by the fast component the emission spectrum is very similar to the CW spectrum in Figure 2(b). Whereas as the time window is moved to longer times after the laser pulse there is a clear shift of the peak of the spectrum to higher energy.

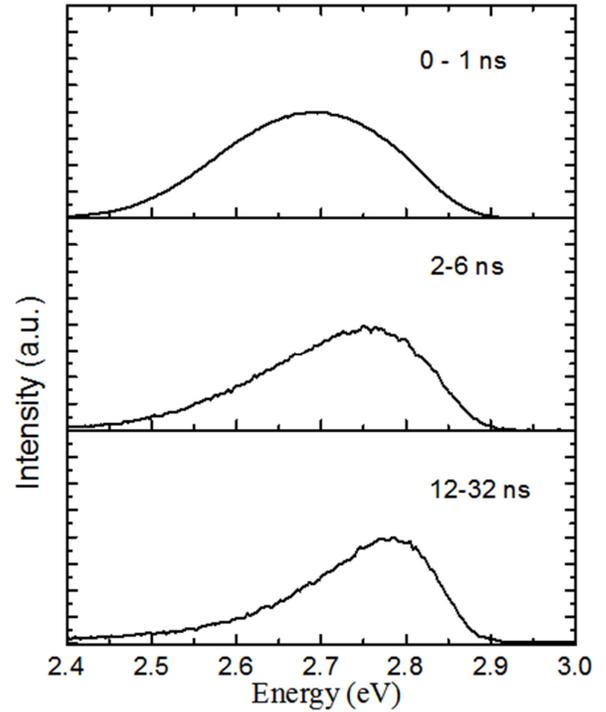

Figure 7. Time resolved spectra for sample A3 for the time windows indicated after the laser pulse which occurred at t = 0.

The nature of the long-lived emission component and its role in reducing the DOLP is far from clear at the moment and we can only speculate about the processes responsible. Clearly, in both the *a*- and *m*-plane samples carrier localisation plays a big role in not only determining the form of the PL spectra but also the recombination energy. This leads us to suggest that in the *a*-plane samples the localisation itself in some parts of the QWs may not only perturb the nature of the localised hole states but also may influence the electron/hole pair configuration reducing the radiative recombination rate. We are currently examining the nanoscale structure of the QWs grown on *a*- and *m*-plane surfaces in some detail to try to elucidate structural differences, which may help to explain the different optical properties of the two types of sample. Structural aspects under investigation include the possibility of non-random indium clustering within the quantum well, as has already been observed for *a*-plane InGaN QWs grown on ELOG substrates using atom probe tomography [31]. Another aspect of interest is the structure of the InGaN/GaN interfaces in terms of both surface step-related to kinetic roughening process and potentially of the formation of three-dimensional quantum dots. Although the data set we are developing is complex and the differences between *a*- and *m*-plane samples are not yet fully understood, initial indications are that for the same indium content, *a*-plane samples are more likely to exhibit non-random indium clustering and have rougher surfaces than *m*-plane samples. Both these factors, and possibly other microstructural issues, could potentially contribute to the explanation of the differences in optical properties between the two families of samples, particularly since it is the *a*-plane materials which deviate most markedly from the expected optical behaviour and also exhibits the greatest microstructural disorder compared to a theoretical quantum well of uniform composition and width. A more detailed microstructural analysis will be the subject of a later publication.

**Summary**

In summary we observed that the maximum DLP of the *a*-plane QW samples is much less than that observed in the *m*-plane samples. Detailed studies of the spectral dependence of the DLP along with time decay measurements and time resolved spectroscopy revealed that the lower DLP of the *a*- plane samples compared with the *m*-plane samples is associated with an additional underlying emission band. As our comparison of the exciton transitions in the PLE spectra with the peak energies of the recombination reveal the effects of very strong localisation we speculate that the underlying emission band in the *a*-plane samples is due to some change in the localisation environment.

**Acknowledgements**
This work was carried out with the support of the United Kingdom Engineering and Physical Sciences Research Council under grant Nos. EP\J001627\1 and EP\J003603\1

References

[1] H. Matsui, N. N. Fellows, S. Nakamura and S. P. Denbaars, *Semicond. Sci. Technol.,* vol. 23, p. 072001, 2008.

[2] P. Waltereit, O. Brandt, A. Trampert, H. T. Grahn, J. Menniger, M. Ramsteiner, M. Reiche and K. H. Ploog, *Nature,* vol. 406, p. 865, 2000.

[3] S. Schulz, T. J. Badcock, M. A. Moram, P. Dawson, M. J. Kappers, C. J. Humphreys and E. P. O'Reilly, *Phys. Rev. B,* vol. 82, p. 125318, 2010.

[4] Q. Sun, C. D. Yerino, Y. Zhang, Y. S. Cho, S.-Y. Kwon, B. H. Kong, H. K. Cho, I.-N. Lee and J. Han, *J. Cryst. Growth,* vol. 311, p. 3824, 2009.

[5] K. Iwata, H. Asahi, K. Asami, R. Kuroiwa and S. Gonda, *Jap. J. Appl. Phys.,* vol. 36, p. L661, 1997.

[6] C. F. Johnson, M. J. Kappers, J. S. Barnard and C. J. Humphreys, *Phys. Stat. Sol. A,* vol. 5, p. 1786, 2008.

[7] Q. S. Paduano, D. W. Weyburne and D. H. Tomich, *J. Cryst. Growth,* vol. 367, p. 104, 2013.

[8] S.-M. Hwang, H. Song, Y. G. Seo, J.-S. Son, J. Kim and K. H. Baik, *Opt. Express,* vol. 19, p. 23036, 2011.

[9] S. H. Park, J. Park, D.-J. You, D. Moon, J. Jang, D.-U. Kim, H. Chang, S. Moon, Y. K. Song, G.-D. Lee, H. Jeon, J. Xu, Y. Nanishi and E. Yoon, *Appl. Phys. Letts.,* vol. 100, p. 191116, 100.

[10] C. Q. Chen, J. W. Yang, H. M. Wang, J. P. Zhang, V. Adivarahan, M. Gaevski, E. Kuokstis, Z. Gong, M. Su and M. A. Kan, *Jap. J. Appl. Phys.,* vol. 42, p. L640, 2003.

[11] T. Guhne, Z. Bougrioua, P. Vennegues, M. Leroux and M. Albrecht, *J. Appl. Phys.,* vol. 101, p. 113101, 2007.

[12] F. Oehler, D. Sutherland, T. Zhu, R. Emery, T. J. Badcock, M. J. Kappers, C. J. Humphreys, P. Dawson and R. A. Oliver, *J. Cryst. Growth,* vol. 48, p. 32, 2014.

[13] S. E. Brinkley, Y.-D. Lin, A. Chakraborty, N. Pfaff, D. Cohen, J. S. Speck, S. Nakamura and S. P. DenBaars, *Appl. Phys. Letts.,* vol. 98, p. 011110, 2011.

[14] Y. G. Seo, K. H. Baik, H. Song, J.-S. Son, K. Oh and S.-M. Hwang, *Optics Express,* vol. 2011, p. 12919, 19.

[15] C. H. Chiu, S. Y. Kuo, M. H. Lo, C. C. Ke, T. C. Wang, Y. T. Lee, H. C. Kuo, T. C. Lu and S. C. Wang, *J. Appl. Phys.,* vol. 105, p. 063105, 2009.

[16] Y. Zhao, R. M. Farell, Y. -R. Wu and J. S. Speck, *Jap. J. Appl. Phys.,* vol. 53, p. 100206, 2014.

[17] R. A. Oliver, F. C.-P. Massabuau, M. J. Kappers, W. A. Phillips, E. J. Thrush, C. C. Tartan, W. E. Blenkhorn, T. J. Badcock, P. Dawson, M. A. Hopkins, D. W. Allsop and C. C. Humphreys, *Appl. Phys. Letts.,* vol. 103, p. 141114, 2013.

[18] M. E. Vickers, M. J. Kappers, T. M. Smeeton, E. J. Thrush, J. S. Barnard and C. J. Humphreys, *J. Appl. Phys.,* vol. 94, p. 1565, 2003.

[19] M. E. Vickers, J. L. Hollander, C. McAleese, M. J. Kappers, M. A. Moram and C. J. Humphreys, *J. Appl. Phys.,* vol. 111, p. 043502, 2012.

[20] D. Kundys, S. Schulz, F. Oehler, D. Sutherland, T. J. Badcock, P. Dawson, M. J. Kappers, R. A. Oliver and C. J. Humphreys, *J. Appl. Phys.,* vol. 115, p. 113106, 2014.

[21] S. Nakagawa, H. Tsujimura, K. Okamoto, M. Kubota and H. Ohta, *Appl. Phys. Letts.,* vol. 91, p. 171110, 2007.

[22] N. F. Gardner, J. C. Kim, J. J. Wierer, Y. C. Shen and M. R. Krames, *Appl. Phys. Leets.,* vol. 86, p. 111101, 2005.

[23] M. Kubota, K. Okamoto, T. Tanaka and H. Ohta, *Appl. Phys. Letts.,* vol. 92, p. 011920, 2008.

[24] D. M. Graham, A. Soltani-Vala, P. Dawson, M. J. Godfrey, T. M. Smeeton, J. S. Barnard, M. J. Kappers and C. J. Humphreys, *J. Appl. Phys.,* vol. 97, p. 103508, 2005.

[25] S. Marcinkevicius, K. M. Kelchner, L. Y. Kuritzky, S. Nakamura, S. P. DenBaars

and J. S. Speck, *Appl. Phys. Letts.,* vol. 103, p. 111107, 2013.

[26] D. Watson-Parris, M. J. Godfrey, P. Dawson, R. A. Oliver, M. J. Galtrey, M. J. Kappers and C. J. Humphreys, *Phys. Rev. B,* vol. 83, p. 115321, 2011.

[27] S. Schulz, M. A. Caro, C. Coughlan and E. P. O'Reilly, *Phys. Rev. B,* vol. 91, p. 035439, 2015.

[28] S. Schulz, D. P. Tanner, E. P. O'Reilly, M. A. Caro, D. Sutherland, M. J. Davies, P. Dawson, F. Tang, J. T. Griffiths, F. Oehler, M. J. Kappers, R. A. Oliver and C. J. Humphreys, "http://arxiv.org/abs/1509.07099," [Online].

[29] J. A. Davidson, P. Dawson, T. Wang, T. Sugahara, J. W. Orton and S. Sakai, *Semicon. Sci. Tech.,* vol. 15, p. 497, 2001.

[30] H. Masui, H. Yamada, K. Iso, S. Nakamura and S. P. DenBaars, *Appl. Phys. Letts.,* vol. 98, p. 011110, 2011.

[31] F. Tang, T. Thu, F. Oehler, W. Y. Fen, J. T. Griffiths, F. C.-P. Massabuau, M. J. Kappers, T. J. Martin, P. A. Bagot, M. P. Moody and R. A. Oliver, *Appl. Phys. Letts.,* vol. 106, p. 072104, 2015.